# Quantifying the Unknown

Maria Grazia Pia, Matej Batic, Marcia Begalli, Anton Lechner, Lina Quintieri and Paolo Saracco

*Abstract*–The issue of how epistemic uncertainties affect the outcome of Monte Carlo simulation is discussed by means of a concrete use case: the simulation of the longitudinal energy deposition profile of low energy protons. A variety of electromagnetic and hadronic physics models is investigated, and their effects are analyzed. Possible systematic effects are highlighted. The results identify requirements for experimental measurements capable of reducing epistemic uncertainties in the simulation.

## I. INTRODUCTION

The investigation and quantification of epistemic uncertainties [1] is well established in the domain of deterministic simulation, but it is a relatively new domain of research within the scope of Monte Carlo simulation. It concerns the issue of how epistemic uncertainties, i.e. uncertainties due to lack of knowledge – namely in modeling physics processes, affect the outcome of Monte Carlo simulation. In this contest the issue of the transformation of epistemic uncertainties into systematic ones is especially important, since they can have a negative effect on the accuracy and reliability of simulation results.

This study assesses the impact of epistemic uncertainties associated with various physics models and parameters relevant to Monte Carlo codes through the simulation of a concrete use case. The outcome associated with the various models subject to investigation is compared by means of rigorous statistical analysis methods to quantitatively estimate the effect of physics-related systematic uncertainties.

The results are discussed in the context of the applicative environment of the simulation and its associated risks. This further analysis shows how systematic effects determined by inadequate physics knowledge carry different weight - and could even vanish - depending on the application environment (e.g. verification, commissioning, treatment planning etc.). The analysis also shows that the extent of the systematic effects generated by epistemic uncertainties in the physic models depends not only on their intrinsic features, but also on the characteristics of the simulation application environment.

## II. SIMULATION MODELS AND THEIR EPISTEMIC UNCERTAINTIES

The knowledge domain pertinent to the problem of simulating proton depth dose profiles has been assessed through a wide survey of atomic parameters, electromagnetic and nuclear models of proton interactions relevant to the energy range of therapeutic applications.

The investigation involves stopping powers (from ICRU's[2] [2] and Ziegler's [3][4] compilations), the water mean ionization potential, nuclear inelastic cross section data [5], pre-equilibrium models [6], evaporation models (Dostrovsky [7], GEM [8] and ABLA [9]), parameterized hadronic interaction models [10], the Liège INUCL [11] intra-nuclear cascade model, the CHIPS [12] (Chiral Invariant Phase Space) model, various nuclear elastic scattering and multiple Coulomb scattering models [13][14].

The epistemic uncertainties associated with these models and parameters derive either from the lack of experimental results to establish their validity, or from the presence of controversial measurements, whose conflicting results hinder the validation of the simulation. In some cases the lack of documentation of the simulation models themselves is a source of epistemic uncertainty: it prevents the assessment of the ground on which the simulation results stand. A particular case is represented by the lack of clear documentation on the calibration of the simulation models, and of the experimental data used for this purpose.

The physics models addressed in this study are investigated through their implementations in the Geant4 [15][16] toolkit; nevertheless, most of these modeling approaches are common to other major Monte Carlo systems as well. The epistemic uncertainties affecting the simulation results are in large part intrinsic to the modeling approaches themselves; however, some of them are specific to their software implementation.

Several Geant4-based simulations of proton therapy set-ups, like [17]-[24] have been shown to produce results in satisfactory agreement with experimental depth dose measurements in various beam line set-ups. Nevertheless, the characteristics of these experimental environments are not optimal from the perspective of assessing the presence of epistemic uncertainties in the simulation models, since their effect is hidden by calibration and normalization procedures usually applied in the simulation of therapeutical proton beam lines.

## III. SENSITIVITY ANALYSIS

The investigation of the effects of epistemic uncertainties is performed by means of a sensitivity analysis similar to the interval analysis often performed in the domain of

M. G. Pia, M. Batic and P. Saracco are with INFN Sezione di Genova, Via Dodecaneso 33, I-16146 Genova, Italy (telephone: +39 010 3536328, e-mail: MariaGrazia.Pia@ge.infn.it).

M. Begalli is with State University of Rio de Janeiro, Rio de Janeiro, Brazil (email: Marcia.Begalli@cern.ch).

A. Lechner is with the Atomic Institute of the Austrian Universities, Vienna University of Technology, Vienna, Austria and CERN, Geneva, Switzerland (email: Anton.Lechner@cern.ch).

L. Quintieri is with INFN Laboratori Nazionali di Frascati, Frascati, Italy (e-mail: Lina.Quintieri@lnf.infn.it).

deterministic simulation. In that case the effects of options (usually parameters) in the simulation, which are affected by epistemic uncertainties, is evaluated by quantifying the variation of the simulation results they determine, when their values vary across the range of their possible values.

In this study the interval analysis is performed in a similar way regarding the epistemic uncertainties associated with numerical values of physical parameters (e.g. mean ionization potentials and stopping powers). The concept of interval analysis is extended to include analyses where, instead of parameters, variants of modeling approaches are available, which are affected by epistemic uncertainties. In this respect Geant4 is a valuable playground for this kind of analysis, since, by its intrinsic nature as a toolkit it allows the evaluation of multiple physics models in the same simulation environment.

For the purpose of the sensitivity analysis a reference configuration of physics models is defined; differences with respect to the simulation results produced through any given configuration should be considered as potential sources of systematic effects in the simulation, whose outcome would be unstable with respect to its physics modeling choices.

The significance of differences observed in the simulation results is quantified by means of statistical tests: non parametric goodness-of-fit tests implemented in the Statistical Toolkit [25][26] (Kolmogorov-Smirnov, Anderson-Darling and Cramer-von Mises) and of the Wald-Wolfowitz test.

TABLE I. REFERENCE CONFIGURATION OF PHYSICS MODELS DEFINED FOR SENSITIVITY ANALYSIS

| Electromagnetic models and parameters | Hadronic models |
|---|---|
| 75 eV water ionization potential | U-elastic scattering |
| ICRU 49 proton stopping powers | Precompound model |
| EEDL-EPDL models for electrons and photons | Default evaporation |
| Standard models for positrons | Wellisch&Axen inelastic cross sections |

The reference configuration is summarized in Table I. It is worthwhile to stress that this selection of models is only motivated by convenience and does not imply that these models are more representative than others.

IV. SIMULATION SET-UP

The problem is examined with the support of a concrete use case: the simulation of the longitudinal depth dose profile in a proton therapy beam line. This use case, concerning a sensitive application domain (medical physics), highlights the role of Monte Carlo simulation uncertainties in the context of risk analysis.

The simulations are performed in a realistic experimental model, which exploits the geometry set-up of a real-life proton therapy beam line [24] available as an example [20] in the Geant4 toolkit. The energy deposition profiles are scored in a sensitive volume, consisting of a 4 cm cube of water placed at the end of the beam line.

The simulation results presented in the following sections derive from a primary proton beam with a Gaussian energy distribution; the events were generated with primary proton mean energy of 63.95 MeV and 300 keV standard deviation. Primary protons loose energy in the transport through the beam line; their energy distribution at the entrance of the sensitive volume is peaked at approximately 60 MeV and is characterized by a long tail extending to low energy.

The results derive from one million primary protons and were produced with Geant4 9.3, unless differently specified.

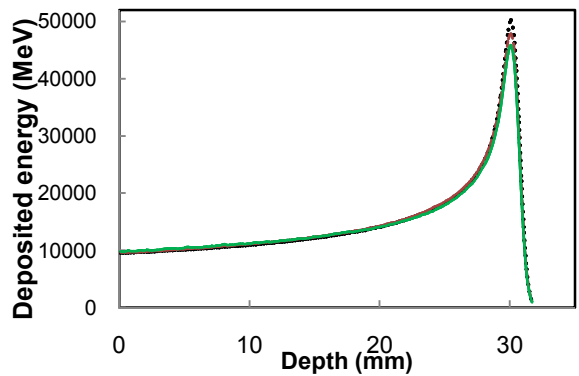

Fig. 1 Longitudinal energy deposition profile resulting from electromagnetic interactions only (black dotted line), electromagnetic and hadronic elastic interactions (dashed red line) and the combination of all of them (solid green line).

Fig. 1 shows the longitudinal energy deposition profile resulting from the effects of electromagnetic interactions only, and of hadronic elastic and inelastic interactions on top of them.

V. RESULTS

Due to space limitations, this paper summarizes a small sample of research results concerning epistemic uncertainties in proton depth dose simulation. A more extensive set of results is documented in another paper [27] along with their quantitative analysis and in-depth discussion.

A. Epistemic uncertainties in electromagnetic models

Epistemic uncertainties affect the value of the mean water ionization potential; values ranging from approximately 61 eV to more than 80 eV are reported in the literature.

TABLE II. MODELS AND PARAMETERS OF PROTON ELECTROMAGNETIC INTERACTIONS INVESTIGATED IN THIS STUDY

| Water ionization potential (eV) | Proton stopping powers |
|---|---|
| 67.2 | ICRU 49 |
| 75 | Ziegler 1977 |
| 80.8 | Ziegler 1985 |
|  | Ziegler 2000 |

The different values of the water mean ionization potential shift the longitudinal position of the Bragg peak by approximately 200 μm with respect to the location associated with the value (75 eV) recommended by ICRU 49 report. The longitudinal energy deposition profiles resulting from the values listed in Table 2 are shown in Fig. 2.

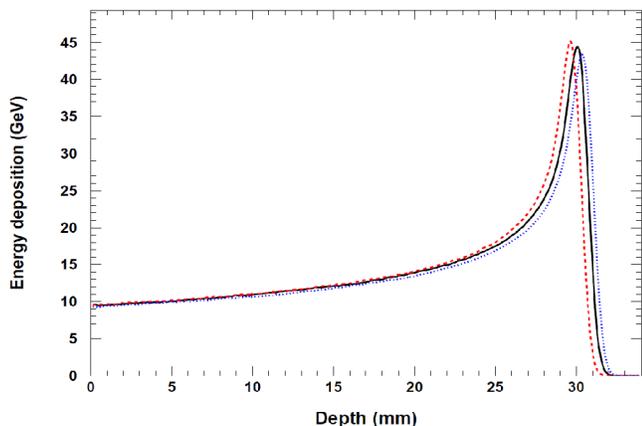

Fig. 2 Longitudinal energy deposition profiles associated with different values of the water mean ionization potential.

Similar shifts are observed when using different proton stopping power compilations, as listed in Table 2. These compilations all derive from fits to experimental data; nevertheless, the lack of consensus in the derivation of stopping power parameterizations from the available data is source of epistemic uncertainties in the simulation results.

The presence of these epistemic uncertainties does not affect the common use of Monte Carlo simulation in proton therapy practice: in that context, where range, rather than the absolute value of the proton beam energy is relevant, it is common practice to calibrate the proton beam parameters (energy and energy spread) to be used in the simulation with respect to experimental data in the same beam line set-up; this adjustment hides any shifts in the Bragg peak location related to epistemic uncertainties in physics models.

Multiple scattering modeling can be source of epistemic uncertainties related to various parameters embedded in models, which govern effects like backscattering and lateral displacement. Some of these settings were investigated through the evolution of the multiple scattering implementation over four Geant4 versions; they are listed in Table 3.

The investigation concerned the generic multiple scattering process G4MultipleScattering available in earlier Geant4 versions and the specialized G4hMultipleScattering process, specific to hadrons, available since Geant4 9.1; these processes were used in their default configuration of models and parameters. An attempt was made to evaluate the effects of the G4WenzelVI model available in Geant4 9.3, which is stated to provide precise simulation for muons and hadrons [28]; it resulted in not letting any primary proton reach the sensitive volume. The probability for observing such a result, given the average acceptance associated with other multiple scattering models, is less than $10^{-6}$. The production with the G4WenzelVI model was limited to a smaller number of events (250000) than the other test cases, since an equivalent simulation sample would have required approximately CPU days, which would have been a prohibitive effort, given the limited computing resources available to the authors.

TABLE III. MULTIPLE SCATTERING MODELS AND PARAMETERS INVESTIGATED IN THIS STUDY

| Geant4 Version | Range Factor | Step Limit | Lateral Displacement |
|---|---|---|---|
| Generic multiple scattering | | | |
| 8.1p02 | 0.02 | 1 | |
| 9.1 | 0.02 | 1 | 1 |
| 9.2p03 | 0.02 | 1 | 1 |
| 9.3 | 0.04 | 1 | 1 |
| Specialized hadron multiple scattering | | | |
| 9.3 | 0.2 | 0 | 1 |

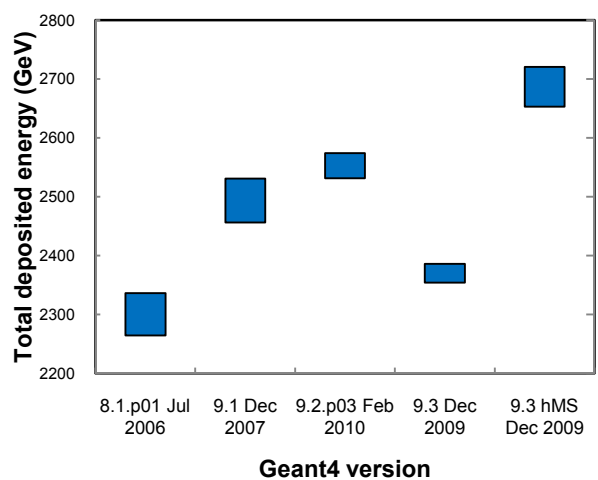

Fig. 3 Total deposited energy in the sensitive volume for various Geant4 versions; the blue bands represent the 95% confidence interval for the mean value calculated over all the hadronic physics configurations considered in the paper; the electromagnetic physics configuration was unchanged in the various test configurations, apart from evolutions in the multiple scattering algorithm.

Due to the scarcity of pertinent experimental data to validate proton multiple scattering models in the energy range relevant to this use case, this simulation domain is characterized by epistemic uncertainties.

The different multiple scattering effects result in significantly large differences in the energy deposited in the sensitive volume, as it can be seen in Fig. 3. These differences do not appear to be related to changes in other simulation modeling domains: in these tests the electromagnetic physics configuration was kept unchanged, apart from the evolutions in the multiple scattering algorithm, and different hadronic models result in consistent outcomes in all Geant4 version; therefore Fig. 3 hints that the incompatibility of the results derives from different multiple scattering settings or other similar issues common to all physics configurations. Plausible correlation of the total energy deposit and the primary proton acceptance is hinted by Fig. 4, which shows the 95% confidence intervals for the mean value of the acceptance (i.e.

the fraction of primary protons reaching the sensitive volume) for various Geant4 versions.

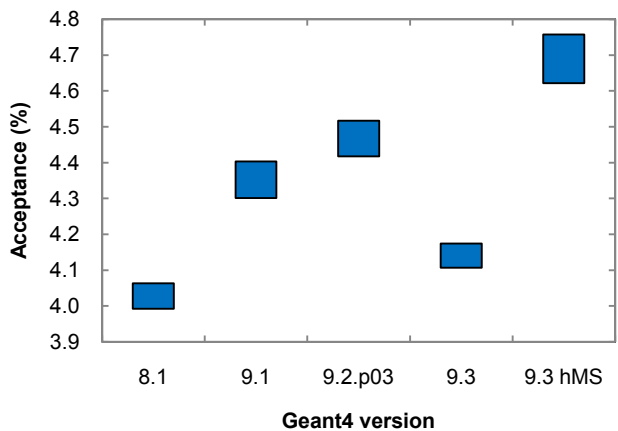

Fig. 4 Primary proton acceptance (i.e. fraction of protons reaching the sensitive volume) for various Geant4 versions; the blue bands represent the 95% confidence interval for the mean value calculated over all the hadronic physics configurations considered in the paper; the electromagnetic physics configuration was unchanged in the various test configurations, apart from evolutions in the multiple scattering algorithm.

The total energy deposited in the sensitive volume appears correlated with the acceptance, i.e. the fraction of protons reaching the sensitive volume after traversing the beam line. Effects of backscattering and lateral displacement related to multiple scattering modeling could be responsible for the observed large differences.

### B. Epistemic uncertainties in hadronic simulation models

Epistemic uncertainties are present in hadronic interaction models due to the limited availability of experimental data for their validation. Moreover, while electromagnetic models are usually fully specified in their formulation, hadronic interaction models are often characterized by a number of parameters, or adjustable modeling options, which are subject to calibration procedures known as "tuning". In general, these procedures are scarcely documented; this situation further contributes to epistemic uncertainties in hadronic simulation models.

The set of hadronic elastic and inelastic models investigated in this study is listed in Table 4. Several of these modeling approaches are common to other Monte Carlo codes, like MCNP, SHIELD-HIT, PHITS and FLUKA, and to GEANT 3.

TABLE IV. HADRONIC ELASTIC AND INELASTIC SCATTERING MODELS INVESTIGATED IN THIS STUDY

| Elastic | Inelastic |
|---|---|
| LEP (parameterized) | LEP (parameterized) |
| U-elastic | Precompound |
| Bertini-elastic | Precompound-GEM |
| CHIPS-elastic | Precompound-Fermi break-up |
|  | Binary cascade |
|  | Bertini cascade |
|  | Liège cascade |
|  | CHIPS-inelastic |

These elastic and inelastic modeling options produce equivalent longitudinal energy deposition profiles; this conclusion is supported by the results of the goodness-of-fit tests, which produce p-values ranging from 0.85 to 1 for all the longitudinal energy deposition profiles subject to comparison.

Nevertheless, small systematic effects in the longitudinal energy deposition profiles resulting from different hadronic simulation models are detected by the Wald-Wolfowitz test. This test is complementary to goodness-of-fit tests, being sensitive to the sign of differences between two distributions, while goodness-of-fit tests are sensitive to their distance. The systematic effects highlighted by this test are smaller than 2%.

Despite the similarity of longitudinal energy deposition profiles, significant differences are visible in the secondary particle production resulting from the various hadronic inelastic models listed in Table 4. An example is shown in Fig. 5 Energy spectrum of secondary neutrons produced with different configurations of the Geant4 Precompound model: with Geant4 default evaporation model (black circles), with GEM evaporation (red squares), activating Fermi break up (blue triangles) and activating the Binary Cascade model (white crosses), which in turn invokes the Precompound model to handle the preequilibrium phase., which plots the energy spectrum of neutrons produced by different simulation models.

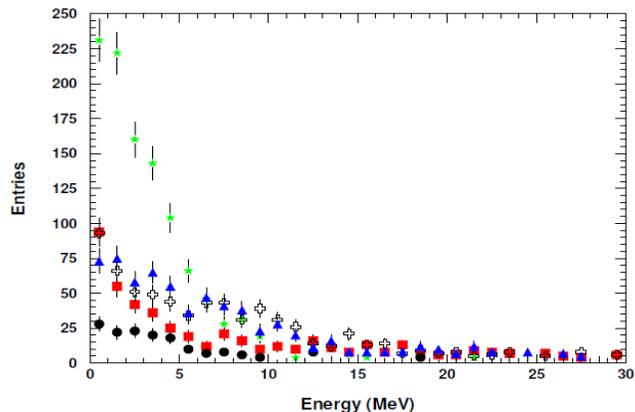

Fig. 5 Energy spectrum of secondary neutrons produced with different configurations of the Geant4 Precompound model: with Geant4 default evaporation model (black circles), with GEM evaporation (red squares), activating Fermi break up (blue triangles) and activating the Binary Cascade model (white crosses), which in turn invokes the Precompound model to handle the preequilibrium phase.

## VI. CONCLUSION

Epistemic uncertainties are present in physics models pertinent to the simulation of proton depth dose; some of them, which broadly represent the variety of approaches to describe proton interactions with matter in the energy range up to approximately 100 MeV, have been evaluated in this study.

Epistemic uncertainties affecting the electromagnetic simulation domain value concern the water mean ionization potential and proton stopping powers; they produce systematic effects on the depth of the Bragg peak.

Epistemic uncertainties in the hadronic domain derive from intrinsic differences in the physics models and the parameters they use, for which limited experimental evidence of their validation is available. The differences, and potential systematic effects, they produce on depth dose profiles are comparable with typical experimental uncertainties in proton therapy practice; larger differences are evident in secondary particle spectra.

The largest effects of physics-related epistemic uncertainties are observed in relation to multiple scattering modeling. However, these effects are relevant only when accurate determination of the absolute dose released to the target is required (for instance, in radiation protection applications). Common practices in radiotherapy applications, like the normalization of the simulated dose to a reference value, would hide the systematic effects deriving from the presence of epistemic uncertainty in multiple scattering modeling.

The quantitative evaluation of systematic effects related to epistemic uncertainties in physics models provides insight for the design of experiments suitable to reduce, or cancel their effects.

Further research on methods to identify and quantify epistemic uncertainties, and to deal with them in Monte Carlo software design, is in progress.

The complete set of results is documented and discussed in depth in a dedicated publication [27].

ACKNOWLEDGMENT

The authors are grateful to CERN for support to the research described in this paper. CERN Library's support has been essential to this study; the authors are especially grateful to Tullio Basaglia.

The authors thank Andreas Pfeiffer for his significant help with data analysis tools throughout the project; Katsuya Amako, Sergio Bertolucci, Luciano Catani, Gloria Corti, Andrea Dotti, Gunter Folger, Simone Giani, Vladimir Grichine, Aatos Heikkinen, Alexander Howard, Vladimir Ivanchenko, Mikhail Kossov, Vicente Lara, Katia Parodi, Alberto Ribon, Takashi Sasaki, Vladimir Uzhinsky and Hans-Peter Wellisch for valuable discussions, and Anita Hollier for proofreading help.

The authors do not intend to express criticism, nor praise regarding any of the Monte Carlo codes mentioned in this paper; the purpose of the paper is limited to documenting technical results.